# Measurement-free reconstruction circuit of quantum secrets in quantum secret sharing

Shogo Chiwaki[1] · Ryutaroh Matsumoto[1]

**Abstract**
For a quantum secret sharing scheme built from a general quantum stabilizer code, no measurement-free circuit has been known for reconstructing its quantum secrets, except particular classes, such as one proposed by Cleve, Gottesman and Lo. We propose a measurement-free reconstruction circuit of quantum secrets in quantum secret sharing based on stabilizer codes. Our reconstruction circuit has width $k+|J|$ and consists of $O(k|J|)$ one- or two-qudit unitary gates when $|J|$ participants reconstruct $k$-qudit quantum secrets.

## 1 Introduction

Secret sharing is a cryptographic protocol to share a secret by multiple participants [5, 33]. In a secret sharing scheme, a secret is encoded by the dealer into multiple pieces of information, called shares, then each share is distributed to each participant from the dealer. When the secret is needed, a set of participants collectively reconstructs the secret. A set of participants that can reconstruct secrets is called qualified, a set of participants that has absolutely no information about secrets is called forbidden, and a set neither qualified or forbidden is called intermediate. Nowadays, secret sharing schemes are used in practice to increase robustness and secrecy of distributed storage systems [3], which store classical information. Quantum secret sharing [11, 20, 23], which accepts quantum secrets, is expected to play a similar role to the classical secret sharing [5, 33], in the future quantum internet era [6, 21]. There are two categories of quantum secret sharing, one is based on quantum error correction [11, 15, 25, 26, 31],

✉ Ryutaroh Matsumoto
ryutaroh@ict.e.titech.ac.jp

1 Department of Information and Communications Engineering, Institute of Science Tokyo, Ookayama 2-12-1, Meguro, Tokyo 152-8550, Japan

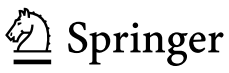

and the other is based on quantum teleportation [20, 23]. In this paper, we focus on the former.

An erasure in quantum and classical error correction means an error whose position in a codeword is known [4, 17]. It is known [11, 15] that a quantum secret sharing scheme can be obtained from any quantum error-correcting code, in which the dealer distribute each quantum symbol in a codeword to each participant as a share, after a given quantum secret is encoded to the codeword by an encoder of the underlying code. A qualified set of participants can reconstruct secrets by an erasure correction procedure of the code by regarding missing shares as erasures in a codeword [11, 15, 27]. The stabilizer code [2, 7, 8, 14, 24] is a class of quantum error-correcting code allowing efficient implementation of encoders and decoders, and in this paper we focus on quantum secret sharing based on stabilizer codes.

The standard and popular way [12] of erasure correction involves measurements. Measurements are costly on some physical devices and measurement-free fault-tolerant computation has been actively investigated recently [19, 30, 34]. Measurement-free reconstruction procedures of quantum secrets are known for specific classes of quantum secret sharing schemes [9, 11, 15, 28, 29, 36], which support limited classes of access structure. An access structure of a secret sharing scheme is a family of qualified sets and a family of forbidden sets, and a desired access structure varies with use cases. Secret sharing schemes based on general stabilizer codes enable much wider classes of access structures [27]. For secret sharing schemes based on general stabilizer codes, it was merely shown that measurement-free reconstruction procedures theoretically exist [27], but explicit and efficient measurement-free reconstruction procedures or circuits are required on physical devices with costly measurements. In this paper we propose such a required circuit. The proposed circuit has width $k + |J|$ and consists of $O(k|J|)$ unitary gates acting on either one or two qudits, when $|J|$ participants reconstruct $k$-qudit quantum secrets. By the width of a quantum circuit, we mean the total number of input qudits and ancillary qudits. Our proposed circuit is the inverse of final steps in the encoding circuit for stabilizer codes proposed in [16, Section 4].

This paper is organized as follows: In Sect. 2, some useful facts in the stabilizer codes are collected and reviewed. Section 3 proposes our measurement-free reconstruction circuit. Sections 2 and 3 have several examples to illustrate results. In Sect. 4, concluding remarks are given. Since we do not propose a new class of quantum secret sharing, we do not have to discuss the security criteria and the parameters (e.g., access structures, coding rates, etc.) of stabilizer-based quantum secret sharing considered in this paper. They can be found in the existing literature [11, 15, 27, 29], and readers are referred to them.

## 2 Some properties of the quantum stabilizer codes

Let $p$ be a prime integer. Denote by $\mathcal{H}_p$ the $p$-dimensional complex linear spaces. A quantum state in $\mathcal{H}_p$ is called a *qudit*. When quantum secrets and shares are quantum states in tensor products of $p^m$-dimensional complex linear spaces, they can also be regarded as tensor products of $\mathcal{H}_p$. So we may assume without much loss of generality that each quantum symbol in secrets and shares belongs to $\mathcal{H}_p$, and we will consider stabilizer codes whose codewords belong to $\mathcal{H}_p^{\otimes n}$.

Let $\mathbf{F}_p = \{0, 1, \ldots, p-1\}$ be the finite field with $p$ elements. We will use the notation of vectors and their associated unitary matrices introduced by Calderbank et al. [8], which is one of the standard notations in expression of quantum stabilizer codes by finite field vectors. For a vector $\boldsymbol{x} = (a_1, \ldots, a_n | b_1, \ldots, b_n) \in \mathbf{F}_p^{2n}$ and $J \subset \{1, \ldots, n\}$, let $P_J(\boldsymbol{x})$ be the projection map from $\mathbf{F}_p^{2n}$ to $\mathbf{F}_p^{2|J|}$, sending $\boldsymbol{x}$ to $(a_{i_1}, \ldots, a_{i_{|J|}} | b_{i_1}, \ldots, b_{i_{|J|}})$, where $\{i_1, \ldots, i_{|J|}\} = J$. By $\mathbf{F}_p^J \subset \mathbf{F}_p^{2n}$ we mean $\{(a_1, \ldots, a_n | b_1, \ldots, b_n) \in \mathbf{F}_p^{2n} : a_i = b_i = 0 \text{ if } i \notin J\}$.

***Example 1*** Let $p = 3$ and $n = 6$, which remain the same in the subsequent examples. Consider $\boldsymbol{x} = (1, 0, 0, 2, 0, 2 | 0, 2, 0, 1, 1, 2) \in \mathbf{F}_3^{12}$. In the subsequent examples, we omit the commas and $\boldsymbol{x}$ will be denoted by (100202|020112). Consider $J = \{3, 4, 5, 6\}$, then its complement $\overline{J}$ becomes $\{1, 2\}$, and we have $i_1 = 3$, $i_2 = 4$, $i_3 = 5$, $i_4 = 6$ in the above definition of $P_J(\boldsymbol{x})$. The projected vector $P_J(\boldsymbol{x})$ is (0202|0112), which has $8 = 2|J|$ numbers in $\mathbf{F}_3$ as its component.

On the other hand, $\mathbf{F}_p^J = \mathbf{F}_3^{\{3,4,5,6\}}$ is a 8-dimensional linear space with $3^8 = 6561$ vectors, whose basis as a linear space is {(1000|0000), (0100|0000), (0010|0000), (0001|0000), (0000|1000), (0000|0100), (0000|0010), (0000|00010)}.

For a vector $\boldsymbol{x} = (a_1, \ldots, a_n | b_1, \ldots, b_n) \in \mathbf{F}_p^{2n}$, we define the associated $p^n \times p^n$ unitary matrix $M(\boldsymbol{x})$ as $X^{a_1} Z^{b_1} \otimes \cdots \otimes X^{a_n} Z^{b_n}$, where $X|i\rangle = |1 + i \bmod p\rangle$ and $Z|i\rangle = \exp(2\pi\sqrt{-1}/p)^i |i\rangle$. We define $\omega = \exp(2\pi\sqrt{-1}/p)$ for $p \geq 3$ and $\omega = \sqrt{-1}$ for $p = 2$. We will consider the error group $E_n = \{\omega^i X^{a_1} Z^{b_1} \otimes \cdots \otimes X^{a_n} Z^{b_n} : a_1, b_1, \ldots, a_n, b_n \in \mathbf{F}_p \text{ and } i \text{ is an integer }\}$.

***Remark 2*** The other definition of stabilizer codes not used in this paper requires every eigenvalue to be $+1$ [14, 24]. A unitary matrix $X^{a_1} Z^{b_1} \otimes \cdots \otimes X^{a_n} Z^{b_n}$ has eigenvalues $\pm\sqrt{-1}$ when it has the odd number of $XZ$ (e.g., $XZ$ with $n = 1$) and $p = 2$. Eigenvalue $+1$ needs the coefficient $\sqrt{-1}$ in $E_n$ for $p = 2$. For $p \geq 3$, every eigenvalue of $X^{a_1} Z^{b_1} \otimes \cdots \otimes X^{a_n} Z^{b_n}$ is a power of $\exp(2\pi\sqrt{-1}/p)$, and this exceptional treatment is unnecessary for $p \geq 3$.

We always consider the standard symplectic inner product $\langle \cdot, \cdot \rangle$ in $\mathbf{F}_p^{2n}$, that is, the symplectic inner product of $(a_1, \ldots, a_n | b_1, \ldots, b_n)$ and $(a_1', \ldots, a_n' | b_1', \ldots, b_n')$ is $\sum_{i=1}^n (a_i b_i' - a_i' b_i)$, and $C^\perp$ denotes the orthogonal space of $C$ in $\mathbf{F}_p^{2n}$ with respect to that inner product. Let $S \subset E_n$ be a commutative subgroup. There exists a common eigenvector $|\varphi\rangle$ such that, for any $U \in S$, $U|\varphi\rangle = \eta(U)|\varphi\rangle$, where $\eta(U)$, not necessarily equal to $+1$, is the eigenvalue of $|\varphi\rangle$ to $U$. Observe that $\eta(U)$ is always a power of $\omega$. The quantum stabilizer code $Q(S) \subset \mathcal{H}_p^{\otimes n}$ with parameter $[[n, k]]_p$ is defined as $\{|\psi\rangle \in \mathcal{H}_p^{\otimes n} : U|\psi\rangle = \eta(U)|\psi\rangle \text{ for all } U \in S\}$ [2, 7, 8]. The normalizer of $S'$ of $S$ in $E_n$ is $S' = \{V \in E_n : UV = VU \text{ for all } U \in S\}$.

Let $C = M^{-1}(S)$. Then $C$ is an $\mathbf{F}_p$-linear space with $\dim C = n - k$, $C^\perp = M^{-1}(S')$, $C^\perp \supset C$, and $\dim C^\perp = n + k$. By abuse of notation, we also write $Q$(the commutative subgroup generated by $\{M(\boldsymbol{x}) : \boldsymbol{x} \in C\}$) as $Q(C)$. There always exists an $\mathbf{F}_p$-linear subspace $C^m \subset C^\perp$ such that $C^m \supset C$ and $C^m = (C^m)^\perp$ [1, Section 20]. We always have $\dim C^m = n$.

Hereafter, suppose that any erasures in $\overline{J} \subset \{1, \dots, n\}$ are correctable by $Q(C)$. We will reconstruct the original quantum message (not the codeword) by using quantum symbols only in $J = \{1, \dots, n\} \setminus \overline{J}$. We collect several useful facts.

**Proposition 3** *1.* [13, Section 4.2] *For any subspace $D \subset \mathbf{F}_p^{2n}$ and any index set $J' \subset \{1, \dots, n\}$, we have $[P_{J'}(D \cap \mathbf{F}_p^{J'})]^\perp = P_{J'}(D^\perp)$.*

2. [27] *Recall that $Q(C)$ can correct any erasures at $\overline{J}$. We have $C \cap \mathbf{F}_p^{\overline{J}} = C^\perp \cap \mathbf{F}_p^{\overline{J}} = C^m \cap \mathbf{F}_p^{\overline{J}}$.*
3. *The first and the second claims imply $P_{\overline{J}}(C) = P_{\overline{J}}(C^\perp) = P_{\overline{J}}(C^m)$.*
4. *For any $\boldsymbol{x} \in C^\perp$, there exists $\boldsymbol{u} \in C$ such that $\boldsymbol{w} = \boldsymbol{x} - \boldsymbol{u} \in C^\perp \cap \mathbf{F}_p^J$.*
5. *For any $\boldsymbol{z} \in C^m$, there exists $\boldsymbol{v} \in C$ such that $\boldsymbol{y} = \boldsymbol{z} - \boldsymbol{v} \in C^m \cap \mathbf{F}_p^J$.*

***Proof*** By Claim 3, there exists $\boldsymbol{u} \in C$ such that $P_{\overline{J}}(\boldsymbol{x}) = P_{\overline{J}}(\boldsymbol{u})$. Then $\boldsymbol{w} = \boldsymbol{x} - \boldsymbol{u}$ belongs to $C^\perp \cap \mathbf{F}_p^J$. Also by Claim 3, there exists $\boldsymbol{v} \in C$ such that $P_{\overline{J}}(\boldsymbol{z}) = P_{\overline{J}}(\boldsymbol{v})$. Then $\boldsymbol{y} = \boldsymbol{z} - \boldsymbol{v}$ belongs to $C^m \cap \mathbf{F}_p^J$. □

***Example 4*** We give an example to illustrate Proposition 3. Assume $p = 3$, and we consider a $[[6, 2, 3]]_3$ stabilizer code [16, Eq. (13)] defined by a self-orthogonal space $C$ generated by four vectors

$$\boldsymbol{h}_1 = (100202|020112), \tag{1}$$
$$\boldsymbol{h}_2 = (010000|001222), \tag{2}$$
$$\boldsymbol{h}_3 = (001200|220201), \tag{3}$$
$$\boldsymbol{h}_4 = (000011|211002). \tag{4}$$

A self-dual space $C^m \supset C$ can be generated [16, Eq. (13)] by Eqs. (1)–(4) and

$$\boldsymbol{z}_1 = (000100|122000), \tag{5}$$
$$\boldsymbol{z}_2 = (000001|221020). \tag{6}$$

A dual-containing space $C^\perp \supset C^m$ can be generated [16, Eq. (13)] by Eqs. (1)–(6), and

$$\boldsymbol{x}_1 = (000000|101100), \tag{7}$$
$$\boldsymbol{x}_2 = (000000|100021). \tag{8}$$

Let $J = \{3, 4, 5, 6\}$ and $\overline{J} = \{1, 2\}$. Since the minimum distance is $3, 4 = n - d + 1$ qudits can reconstruct secrets. Then we have $C \cap \mathbf{F}_3^{\{1,2\}} = C^m \cap \mathbf{F}_3^{\{1,2\}} = C^\perp \cap \mathbf{F}_3^{\{1,2\}} = \{\mathbf{0}\}$, exemplifying Claim 2 in Proposition 3.

We have

$$\boldsymbol{w}_1 = \boldsymbol{x}_1 - \boldsymbol{h}_3 - \boldsymbol{h}_4 = (002122|000200) \in C^\perp \cap \mathbf{F}_3^{\{3,4,5,6\}}, \tag{9}$$
$$\boldsymbol{w}_2 = \boldsymbol{x}_2 - \boldsymbol{h}_3 - \boldsymbol{h}_4 = (002122|002121) \in C^\perp \cap \mathbf{F}_3^{\{3,4,5,6\}}, \tag{10}$$

exemplifying Claim 4 in Proposition 3. Similarly, we have

$$\boldsymbol{y}_1 = \boldsymbol{z}_1 + \boldsymbol{h}_4 = (000111|000002) \in C^m \cap \mathbf{F}_3^{\{3,4,5,6\}}, \tag{11}$$

$$\boldsymbol{y}_2 = \boldsymbol{z}_2 - \boldsymbol{h}_3 = (002101|001122) \in C^m \cap \mathbf{F}_3^{\{3,4,5,6\}}, \tag{12}$$

exemplifying Claim 5 in Proposition 3.

## 3 Proposed circuit of reconstructing quantum secrets

Suppose that a quantum secret

$$|\boldsymbol{i}\rangle = |i_1\rangle|i_2\rangle\cdots|i_k\rangle \in \mathcal{H}_p^{\otimes k} \tag{13}$$

is given to the dealer, where $\boldsymbol{i} = (i_1, \ldots, i_k) \in \mathbf{F}_p^k$. Then the dealer uses some encoding circuit to encode the quantum message $|\boldsymbol{i}\rangle$ to a codeword $|\overline{\boldsymbol{i}}\rangle$. We assume that the codewords can be expressed as

$$|\overline{\boldsymbol{i}}\rangle = (\alpha_1 M(\boldsymbol{x}_1))^{i_1}\cdots(\alpha_k M(\boldsymbol{x}_k))^{i_k}|\overline{\mathbf{0}}\rangle, \tag{14}$$

where

1. $|\overline{\mathbf{0}}\rangle$ is an eigenvector of every matrix in $M(C^m)$,
2. $\boldsymbol{x}_1 + C^m, \ldots, \boldsymbol{x}_k + C^m$ form a basis of the quotient linear space $C^\perp/C^m$, $\alpha_1, \ldots, \alpha_k$ are powers of $\omega$,
3. there exist $\boldsymbol{z}_i \in C^m$ such that $\langle \boldsymbol{x}_i, \boldsymbol{z}_j\rangle = \delta_{i,j}$ for $i = 1, \ldots, k$,
4. and $|\overline{\mathbf{0}}\rangle$ belongs to eigenvalue $+1$ of $\alpha_i^{-1} M(\boldsymbol{z}_i)$ for $i = 1, \ldots, k$.

For a given stabilizer $S \subset E_n$, one can easily find logical $Z$ operators $\overline{Z}_1, \ldots, \overline{Z}_k$ and logical $X$ operators $\overline{X}_1, \ldots, \overline{X}_k$. Since the logical operators belong to $E_n$, one can express them as $\overline{X}_i = \alpha_i M(\boldsymbol{x}_i)$ and $\overline{Z}_i = \xi_i M(\boldsymbol{x}_i)$ for $i = 1, \ldots, k$, where $\alpha_i$ and $\xi_i$ are powers of $\omega$. By the anti-commuting relation $\overline{X}_i\overline{Z}_i = \exp(-2\pi\sqrt{-1}/p)\overline{Z}_i\overline{X}_i$, it follows that $\xi_i = 1/\alpha_i$. Therefore, when encoding is performed by logical $X$ and $Z$ operators, the above assumptions automatically become true. In particular the assumption (14) is satisfied when the dealer uses an encoding circuit in [10, 16, 18, 32].

***Remark 5*** Assume $p = 2$. As stated in Remark 2, $M(\boldsymbol{z}_i)$ can have eigenvalues $\pm\sqrt{-1}$. So the coefficient $\pm\sqrt{-1}$ is required with $M(\boldsymbol{z}_i)$ having eigenvalues $\pm\sqrt{-1}$ unless $\boldsymbol{z}_i$ is adjusted to eliminate $XZ$ from $M(\boldsymbol{z}_i)$, which in turn requires the coefficient $\alpha_i (\neq 1)$ for the corresponding $\boldsymbol{x}_i$ in Eq. (14).

In his paper [16, Section 4], Grassl proposed an encoding circuit involving $(n+k)$ qudits for stabilizer codes, which proceeds with quantum message $|\boldsymbol{i}\rangle \in \mathcal{H}_p^{\otimes n}$ as follows:

1. The logical zero codeword $|\overline{\mathbf{0}}\rangle$ is prepared.

2. By application of the controlled-$(\alpha_i M(\boldsymbol{x}_i))$ gates for $i = 1, \ldots, k$, $|\boldsymbol{i}\rangle \otimes |\overline{\boldsymbol{0}}\rangle$ is transformed to $|\boldsymbol{i}\rangle \otimes |\overline{\boldsymbol{i}}\rangle$.
3. By application of the inverse quantum Fourier transform once and the controlled-$(\alpha_i^{-1} M(\boldsymbol{z}_i))$ gates for $i = 1, \ldots k$, the state is further transformed to $|\psi_0\rangle \otimes |\overline{\boldsymbol{i}}\rangle$, where $|\psi_0\rangle = \frac{1}{\sqrt{p^k}} \sum_{\boldsymbol{j} \in \mathbf{F}_p^k} |\boldsymbol{j}\rangle$.

Observe that $\alpha_i M(\boldsymbol{x}_i)$ and $\alpha_i^{-1} M(\boldsymbol{z}_i)$ are the logical $X$ and $Z$ gates [16, 18] on the $i$-th logical qudit for $i = 1, \ldots, k$, as we have the relations

$$\alpha_i^{-1} M(\boldsymbol{z}_i)|\overline{\boldsymbol{0}}\rangle = (+1)|\overline{\boldsymbol{0}}\rangle, \tag{15}$$

$$\alpha_i M(\boldsymbol{x}_i)\alpha_i^{-1} M(\boldsymbol{z}_i) = \exp(-2\pi\sqrt{-1}/p)\alpha_i^{-1} M(\boldsymbol{z}_i)\alpha_i M(\boldsymbol{x}_i), \tag{16}$$

for $i = 1, \ldots, k$, which correspond to the relations $Z|0\rangle = (+1)|0\rangle$ and $XZ = \exp(-2\pi\sqrt{-1}/p)ZX$.

By Proposition 3, for each $\boldsymbol{x}_i$ used in the dealer's encoding (14), there exists $\boldsymbol{u}_i \in C$ such that $\boldsymbol{w}_i = \boldsymbol{x}_i - \boldsymbol{u}_i$ belongs to $C^\perp \cap \mathbf{F}_p^J$. For any stabilizer codeword $|\varphi\rangle \in Q(C)$, we have

$$\begin{aligned} M(\boldsymbol{x}_i)|\varphi\rangle &= M(\boldsymbol{w}_i + \boldsymbol{u}_i)|\varphi\rangle && (17)\\ &= M(\boldsymbol{w}_i)\beta_i M(\boldsymbol{u}_i)|\varphi\rangle && (18)\\ &= M(\boldsymbol{w}_i)\beta_i \eta(M(\boldsymbol{u}_i))|\varphi\rangle, && (19) \end{aligned}$$

where $\beta_i$ is a power of $\omega$ dependent on $\boldsymbol{w}_i$ and $\boldsymbol{u}_i$, and independent of $|\varphi\rangle$. Then, Step 2 of the dealer's encoding procedure is equivalent to $k$ applications of the controlled-$(\alpha_i \beta_i \eta(M(\boldsymbol{u}_i))M(\boldsymbol{w}_i))$ gates for $i = 1, \ldots, k$. Observe that the controlled-$(\alpha_i \beta_i \eta(M(\boldsymbol{u}_i))M(\boldsymbol{w}_i))$ gates can be applied by only the participants in $J$ without help of those in $\overline{J}$. In addition, the controlled-$(\alpha_i \beta_i \eta(M(\boldsymbol{u}_i))M(\boldsymbol{w}_i))$ is equivalent to application of $P^{e(\alpha_i \beta_i \eta(M(\boldsymbol{u}_i)))}$ to the control qudit before controlled-$M(\boldsymbol{w}_i)$, where $\omega^{e(\alpha_i \beta_i \eta(M(\boldsymbol{u}_i)))} = \alpha_i \beta_i \eta(M(\boldsymbol{u}_i))$, $P = Z$ for $p \geq 3$, and $P = \sqrt{Z}$ for $p = 2$.

In a similar way, by Proposition 3, for each $\boldsymbol{z}_i$ used in the dealer's encoding (14), there exists $\boldsymbol{v}_i \in C$ such that $\boldsymbol{y}_i = \boldsymbol{z}_i - \boldsymbol{v}_i$ belongs to $C^m \cap \mathbf{F}_p^J$. For any stabilizer codeword $|\varphi\rangle \in Q(C)$, we have

$$\begin{aligned} M(\boldsymbol{z}_i)|\varphi\rangle &= M(\boldsymbol{y}_i + \boldsymbol{v}_i)|\varphi\rangle && (20)\\ &= M(\boldsymbol{y}_i)\gamma_i M(\boldsymbol{v}_i)|\varphi\rangle && (21)\\ &= M(\boldsymbol{y}_i)\gamma_i \eta(M(\boldsymbol{v}_i))|\varphi\rangle, && (22) \end{aligned}$$

where $\gamma_i$ is a power of $\omega$ dependent on $\boldsymbol{y}_i$ and $\boldsymbol{v}_i$ and independent of $|\varphi\rangle$. Then, Step 3 of the dealer's encoding procedure is equivalent to $k$ applications of the controlled-$(\alpha_i^{-1} \gamma_i \eta(M(\boldsymbol{v}_i))M(\boldsymbol{y}_i))$ gates for $i = 1, \ldots, k$. Observe that the controlled-$(\alpha_i^{-1} \gamma_i \eta(M(\boldsymbol{v}_i))M(\boldsymbol{y}_i))$ gates can be applied by only the participants in $J$ without help of those in $\overline{J}$. In addition, the controlled-$(\alpha_i^{-1} \gamma_i \eta(M(\boldsymbol{v}_i))M(\boldsymbol{y}_i))$ is equivalent to application of $P^{e(\alpha_i^{-1} \gamma_i \eta(M(\boldsymbol{v}_i)))}$ to the control qudit before controlled-$M(\boldsymbol{y}_i)$.

Let $\rho_J$ be the density matrix of quantum shares in $J$. Observe that, because shares in $\overline{J}$ are missing, $\rho_J$ is almost always a mixed state. The participants in a qualified set $J$ can reconstruct the quantum secret as follows:

1. Prepare $|\psi_0\rangle = \frac{1}{\sqrt{p^k}} \sum_{\boldsymbol{j} \in \mathbf{F}_p^k} |\boldsymbol{j}\rangle$.
2. Apply the inverse of the controlled-$M(\boldsymbol{y}_i)$ with the $i$-th qudit in $|\psi_0\rangle$ as its control and $\rho_J$ as its target, for $i = 1, \ldots, k$.
3. Apply $P^{-e(\alpha_i^{-1}\gamma_i\eta(M(\boldsymbol{v}_i)))}$ to the $i$-th qudit in $|\psi_0\rangle$.
4. Apply the quantum Fourier transform to the result of $|\psi_0\rangle$.
5. Apply the inverse of the controlled-$M(\boldsymbol{w}_i)$ with the $i$-th qudit in the result of the quantum Fourier transform as its control and the result of $\rho_J$ as its target, for $i = 1$, $\ldots$, $k$.
6. Apply $P^{-e(\alpha_i\beta_i\eta(M(\boldsymbol{u}_i)))}$ to the $i$-th qudit in $|\psi_0\rangle$.

Since the above steps are the inverse of Steps 2 and 3 by the dealer, the quantum secret is restored in the control part with no entanglement with any other systems.

Since the controlled-$M(\boldsymbol{w}_i)$ and the controlled-$M(\boldsymbol{y}_i)$ can be realized $|J|$ applications of two-qudit controlling gates, the proposed reconstruction circuit contains at most $2k|J|$ two-qudit controlling gates in Steps 2 and 5 with depth at most $2k|J|$ in total, at most $2k$ one-qudit gates in Steps 3 and 6 with depth at most 2 in total, and $O(k^2)$ gates for the quantum Fourier transform in Step 4. Since $k \leq |J|$, the proposed reconstruction contains $O(k|J|)$ gates in total, including the quantum Fourier transform in Step 4. The depth is at most $2(1 + k|J|)$ plus that of the quantum Fourier transform on $k$ qudits. The set of used gates are the same as [16, 18], and decomposition of them into more elementary ones are discussed there. So far, we have not found any variation of the proposed circuit to adjust the trade-off between depth versus width.

Figure 1 shows an example of the proposed reconstruction circuit for the secret sharing scheme based on the $[[6, 2, 3]]_3$ stabilizer code and $J = \{3, 4, 5, 6\}$.

***Example 6*** In this example, we give explicit values for variables in the proposed reconstruction of secrets. We retain notations from Example 4. Because of $p = 3$, by Remarks 2 and 5 we can assume the logical zero state $|\overline{\mathbf{0}}\rangle$ belongs to eigenvalue $+1$ of $M(\boldsymbol{h}_i)$ for $i = 1, \ldots, 4$ and of $M(\boldsymbol{z}_j)$ for $j = 1, 2$, and we can also choose $\alpha_1 = \alpha_2 = 1$. From Example 4, we have

$$\boldsymbol{u}_1 = \boldsymbol{u}_2 = \boldsymbol{h}_3 + \boldsymbol{h}_4 = (001211|101200), \tag{23}$$

$$\boldsymbol{v}_1 = -\boldsymbol{h}_4 = (000022|122001), \tag{24}$$

$$\boldsymbol{v}_2 = \boldsymbol{h}_3 = (001200|220201). \tag{25}$$

Then we have

$$M(\boldsymbol{w}_1)M(\boldsymbol{u}_1) \tag{26}$$

$$= (I \otimes I \otimes X^2 \otimes XZ^2 \otimes X^2 \otimes X^2)(Z \otimes I \otimes XZ \otimes X^2Z^2 \otimes X \otimes X) \tag{27}$$

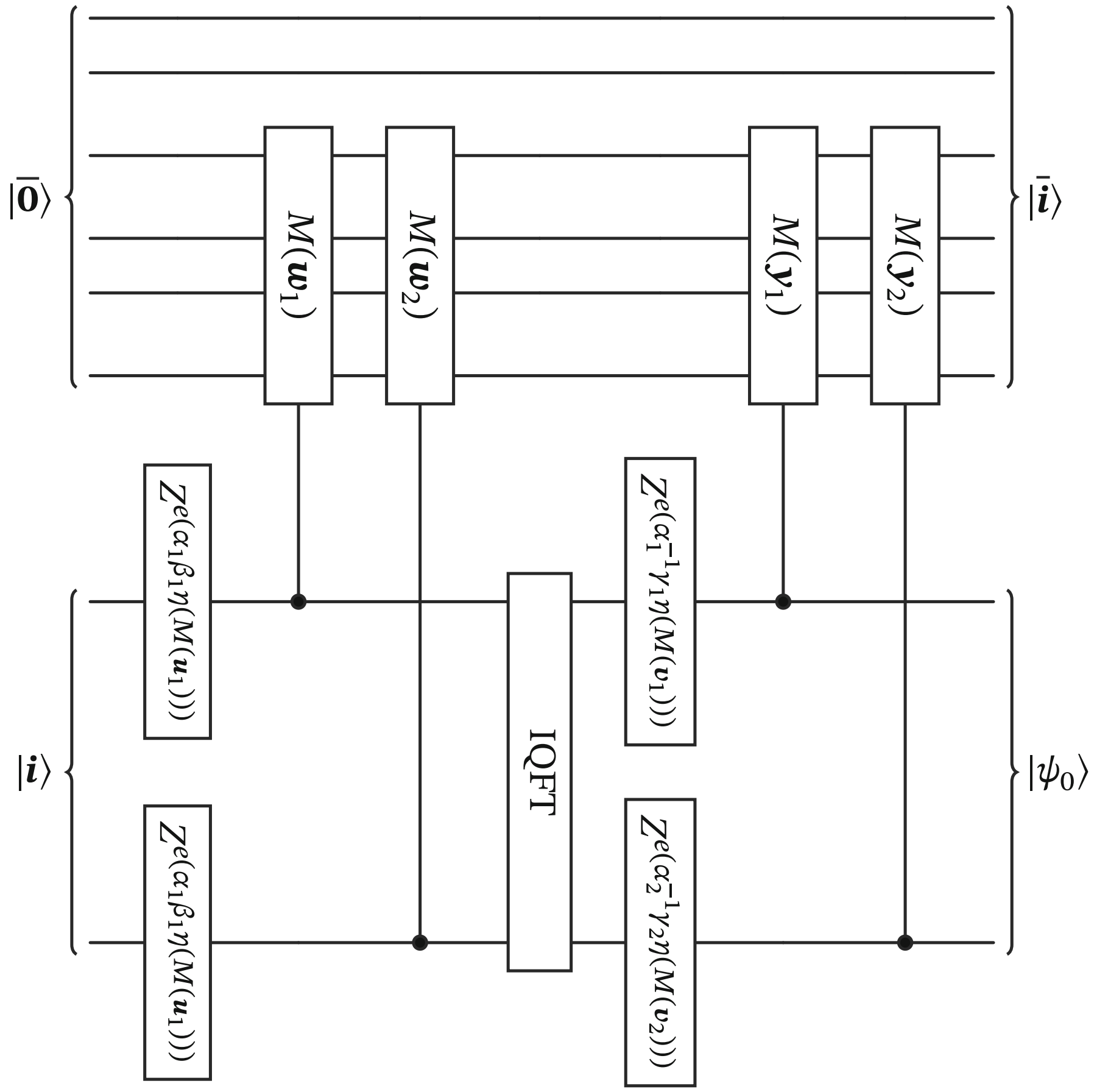

**Fig. 1** The proposed reconstruction circuit for the secret sharing scheme based on the $[[6, 2, 3]]_3$ stabilizer code by the 3rd to the 6th participants, corresponding to Examples 4 and 6. When the above circuit is read from left to right, an encoding circuit for the $[[6, 2, 3]]_3$ stabilizer code is obtained with the logical zero codeword $|\bar{\mathbf{0}}\rangle$ and a message (secret) $|\boldsymbol{i}\rangle$ as its input. The proposed reconstruction circuit can be obtained by reading the above circuit from right to left. The width is $k + n = 8$, which corresponds to the 8 rows in the figure. The 1st to the 6th rows correspond to the 1st to the 6th shares, and the 7th and 8th rows correspond to the 2-qudit quantum secret. Observe that there is no operation on the 1st or the 2nd share, and the net width is $8 - 2 = 6 = k + |J|$

$$= Z \otimes I \otimes Z \otimes \underbrace{XZ^2X^2Z^2}_{=\omega^2 Z} \otimes I \otimes I \tag{28}$$

$$= \omega^2 M(\boldsymbol{x}_1), \tag{29}$$

and we find $\beta_1 = \omega^2$. Similarly, we can find $\beta_2, \gamma_1, \gamma_2$.

We will find $\eta(M(\boldsymbol{u}_i))$ $(i = 1, 2)$. For any stabilizer codeword $|\varphi\rangle \in Q(C)$ and $i = 1, 2$, we have

$$
\begin{aligned}
|\varphi\rangle &= M(\boldsymbol{h}_3)M(\boldsymbol{h}_4)|\varphi\rangle && (30)\\
&= (Z^2 \otimes Z^2 \otimes X \otimes X^2Z^2 \otimes I \otimes Z)(Z^2 \otimes Z \otimes Z \otimes I \otimes X \otimes XZ^2)|\varphi\rangle && (31)\\
&= (Z \otimes I \otimes XZ \otimes X^2Z^2 \otimes X \otimes \underbrace{ZXZ^2}_{=\omega X})|\varphi\rangle && (32)\\
&= \omega M(\boldsymbol{u}_i)|\varphi\rangle, && (33)
\end{aligned}
$$

which implies $M(\boldsymbol{u}_i)|\varphi\rangle = \omega^{-1}|\varphi\rangle$. This means that $\eta(M(\boldsymbol{u}_i)) = \omega^{-1}$. Since $\omega^p = \omega^3 = +1$, we have $\eta(M(\boldsymbol{u}_i)) = \omega^2 \neq +1$. Similarly we can find $\eta(M(\boldsymbol{v}_i)) = +1$ for $i = 1, 2$.

Observe that

- $\beta_i$, $\omega_i$, $\eta(M(\boldsymbol{u}_i))$ and $\eta(M(\boldsymbol{v}_i))$ can be $\neq +1$ even when the eigenvalue $+1$ is required in the definition of stabilizer codes as [14, 24], and,
- even when $\alpha_i$ in Eq. (14) and eigenvalues in the stabilizer code definition are chosen to be $+1$, Steps 3 and 6 cannot be removed in general from our proposed reconstruction.

## 4 Concluding remarks

In this paper, we considered stabilizer-based quantum secret sharing, in which the dealer encodes a $k$-qudit quantum secret into a stabilizer codeword, each quantum symbol in the codeword is distributed to each participant as a quantum share, and finally a qualified set $J$ of participants collectively reconstructs the quantum secret from available shares. We propose a measurement-free reconstruction circuit suitable on physical devices with costly measurements. Our circuit has width $k + |J|$ and consists of $O(k|J|)$ unitary gates, including the quantum Fourier transform on $k$ qudits. Its depth is at most $2(1 + k|J|)$ without counting the quantum Fourier transform on $k$ qudits. When one implements the proposed method with current quantum computers, $p$ is only allowed to be 2, as they are composed of qubits and not of qudits ($p \geq 3$). For $p = 2$, the set of used gates is a subset of those in [10, 16, 18, 32], and the difficulty of practical implementation of the proposed method is at most as difficult as [10, 16, 18, 32]. Concretely, the set of gates for $p = 2$ consists of the controlled-$X$, the controlled-$Z$, and powers of the phase gate $\sqrt{Z}$, in addition to the gates used in the quantum Fourier transform on $k$ qudits. Implementation of such basic gates should not be much difficult.

On the other hand, the erasure correction also attracts attention in the fault-tolerant computation [22, 35]. Since our proposed reconstruction circuit can reconstruct quantum messages encoded in stabilizer codewords with erasures, it is natural to ask if our proposal can be used for erasure correction in the fault-tolerant computation. In the fault-tolerant computation, reconstruction of codewords is required, in contrast to the problem considered in this paper.

Our proposed circuit can reconstruct quantum messages from stabilizer codewords with erasures, then the codewords can be reconstructed by encoding the quantum messages. However, this method of reconstructing codewords is not fault-tolerant, as single error or erasure can propagate to multiple ones during reconstruction. It is a future research agenda to find fault-tolerant reconstruction of codewords based on our proposal.

**Acknowledgements** This work is in part supported by the Japan Society for Promotion of Science under Grant No. 23K10980. The authors would like to thank anonymous reviewers whose comments improved this paper during review process.

**Author Contributions** S.C. formulated the research problem and provided the essential idea, Proposition 2 and uses of eigenvalues and relative phases of quantum states, for solving the problem. R.M. provided the rest of ideas and wrote the manuscript.

**Data availability** No datasets were generated or analyzed during the current study.

## Declarations

**Conflict of interest** The authors declare no Conflict of interest.